\documentclass[twocolumn, aps, prb]{revtex4}
\usepackage{graphicx}
\usepackage{amssymb}
\usepackage{amsmath}

\def\Journal #1,#2,#3,#4#5#6#7{#1 {\bf #2} (#4#5#6#7) #3}
\def\i{\/\/{{\rm i}}\/\/}
\def\e{\/\/{{\rm e}}\/\/}
\def\d{\/\/{{\rm d}}}
\def\lsim{\lower -0.3ex \hbox{$<$} \kern -0.75em \lower 0.7ex \hbox{$\sim$}}
\def\gsim{\lower -0.3ex \hbox{$>$} \kern -0.75em \lower 0.7ex \hbox{$\sim$}}

\begin{document}

\title{Optical response of finite-length carbon nanotubes}
\author{Takeshi Nakanishi and Tsuneya Ando$^1$}
\affiliation{
Nanotube Research Center, AIST
 1--1--1 Higashi, Tsukuba 305--8565, Japan}
\affiliation{
$^1$ Department of Physics, Tokyo Institute of Technology
2--12--1 Ookayama, Meguro-ku, Tokyo 152--8551, Japan}
\date{\today}

\begin{abstract}
Optical response of finite-length metallic carbon nanotubes is calculated including effects of induced edge charges in a self-consistent manner.
The results show that the main resonance corresponding to excitation of the fundamental plasmon mode with wave vector $\pi/l$ with $l$ being the tube length is quite robust and unaffected.
This arises because the strong electric field associated with edge charges is screened and decays rapidly inside the nanotube.
For higher-frequency resonances, the field starts to be mixed and tends to shift resonances to higher frequencies.
\end{abstract}

\maketitle

\section{Introduction}

%
Carbon nanotubes are a one-dimensional conductor consisting of rolled graphite sheets.\cite{Iijima 1991a}
Usually a single-wall carbon nanotube has a diameter of the order of a few nm and a length of the order of several $\mu$m.
Recently, cutting and length selection of single-wall carbon nanotubes became possible.\cite{Duesberg et al 1998a,O'Connell et al 2001a,Chen et al 2001a,Ziegler et al 2005a,Ziegler et al 2005b,Arnold et al 2006a,Hennrich et al 2007a,Fagan et al 2008a,Sun et al 2008a}
In short nanotubes with good conductivity, edge effects start to play important roles in their electronic properties.
The purpose of this paper is to theoretically study low-frequency optical response in metallic carbon nanotubes with finite length.
\par
%
In optical response, induced charges associated with optical transitions often play important roles.
For example, effects of an electric field induced by the polarization of nanotubes are quite significant and tend to suppress interband transitions for light with electric-field polarization perpendicular to the axis.\cite{Ajiki and Ando 1994a,Ajiki and Ando 1995d}
However, strong exciton effects due to the one-dimensional feature\cite{Ando 1997a,Ando 2004c} cause reappearance of exciton peaks even in the presence of the strong depolarization effect.\cite{Uryu and Ando 2006c,Uryu and Ando 2007c}
\par
%
For electric field parallel to the axis, this depolarization effect does not play a role and the absorption is described by the usual dynamical conductivity.
The dynamical conductivity in metallic nanotubes in the low-frequency region was studied\cite{Ando 2002d,Asada and Ando 2006a} in relation to the absence of backscattering\cite{Ando and Nakanishi 1998a,Ando et al 1998b,Nakanishi and Ando 1999a} and the presence of a perfect channel.\cite{Ando and Suzuura 2002a}
In optical response of finite-length carbon nanotubes, often called antenna effects, accumulated charges at both ends of nanotubes can become important even in field parallel to the axis.
\par
%
There have been several theoretical calculations on light scattering by finite-length nanotubes in low-frequency region.
Direct analysis based on the integral equations of electromagnetics was reported.\cite{Slepyan et al 1999a,Slepyan et al 2001a,Slepyan et al 2006a}
Numerical studies were reported on electromagnetic waves in a planar nanotube array.\cite{Hanson 2005a,Hao and Hanson 2006a}
An equivalent circuit model with quantum capacitance and kinetic inductance was considered.\cite{Burke et al 2006a}
\par
%
Recently, nanoscale antenna operation of a carbon nanotube array was experimentally demonstrated.\cite{Wang et al 2004b}
Optical absorption in low-frequency ($\sim$THz) region was reported for nanotube bundles, where a sharp absorption peak was observed in aligned nanotubes but not in sprayed samples.\cite{Akima et al 2006a}
A broad absorption peak was also reported, but attributed to a narrow gap in quasi-metallic or narrow-gap nanotubes.\cite{Ugawa et al 1999a,Itkis et al 2002a,Kampfrath et al 2008a}
Finite-length effects have been considered in connection with various phenomena such as Raman spectroscopy,\cite{Jorio et al 2002c,Dresselhaus 2004a} $\pi$ plasmon absorpt\-ion,\cite{Murakami et al 2005b} etc.
\par
%
In this paper, we shall calculate optical response of finite-length metallic nanotubes in much lower frequency region than an inter-band transition.
In \S2, the method to calculate the optical response is described and an approximation based on excitation of a single plasmon mode in an infinitely long nanotube is introduced.
Numerical results are presented in \S3.
A discussion and summary are given in \S4.
\par
%
\section{Formulation}
%
\subsection{Optical Response}
%
We consider a carbon nanotube with a finite length $l$, lying along the $y$ direction in the range $\!-\!l/2\!<\!y\!<\!+\!l/2$.
Let $E_{\rm ext}(y)\e^{-\i\omega t}$ be an external electric field of incident light and $E(y,\omega)\e^{-\i\omega t}$ be the effective electric field including effects of polarization charges.
The response of the system can generally be described by a nonlocal conductivity $\sigma(y,y')$.
\par
%
Then, the induced current $j(y)\e^{-\i\omega t}$ is given by
%
\begin{equation}
j(y) = \int \! \sigma(y,y')E(y') \, \d y' .
\end{equation}
%
The corresponding induced charge $\rho(y)\e^{-\i\omega t}$ is determined by the equation of continuity:
%
\begin{equation}
{\partial \rho \over \partial t } + {\partial j \over \partial y} = 0 ,
\end{equation}
%
as
%
\begin{equation}
\rho(y) = {1 \over \i \omega} {\partial \over \partial y} j(y) = {1 \over \i \omega} {\partial \over \partial y} \int \! \d y' \, \sigma(y,y') E(y') .
\end{equation}
%
The corresponding scalar potential $\phi(y)\e^{-\i\omega t}$ becomes
%
\begin{equation}
\phi(y) = \int K(y\!-\!y') \rho(y') \d y' ,
\end{equation}
%
where $K(y)$ is the kernel of the Coulomb interaction for cylindrical charge distribution,\cite{Ando 1997a,Wang et al 1992a,Lin and Shung 1993a,Lin and Shung 1993b,Sato et al 1993a,Lin-Chung and Rajagopal 1994a,Gervasoni and Arista 2003a} given by
%
\begin{equation}
K(y) = \int {\d q \over 2\pi} K(q) \, \e^{\i q y } ,
\label{Kernel}
\end{equation}
%
with
%
\begin{equation}
K(q) = {2 \over \kappa} I_0(|q|R) K_0(|q|R) ,
\end{equation}
%
where $R$ is the diameter of the nanotube, $\kappa$ is the static dielectric constant of the environment, and $I_n(t)$ and $K_n(t)$ are the modified Bessel function of the first and second kind, respectively.
The total electric field becomes
%
\begin{equation}
E(y) = E_{\rm ext}(y) + {1 \over \i \omega} \int \!\! \d y' \int \!\! \d y'' \, K'(y\!-\!y') \sigma(y',y'') E(y'') ,
\label{Total Field: Integral Equation}
\end{equation}
%
with
%
\begin{equation}
K'(y\!-\!y') = {\partial^2 \over \partial y \partial y'} K(y\!-\!y') = \int \! {\d q \over 2\pi } K'(q) \e^{\i q ( y-y')} ,
\end{equation}
%
where
%
\begin{equation}
K'(q) = K(q) q^2 .
\end{equation}
%
The power absorption is given by
%
\begin{eqnarray}
P &= &{1\over 2} {\rm Re} \int \! \d y \, j(y) E(y)^* \nonumber\\
&= &{1\over 2} {\rm Re} \int \! \d y \int \! \d y' \, E(y)^* \sigma(y,y') E(y') . 
\label{Power Absorption}
\end{eqnarray}
%
\par
%
We consider the case that the length of the nanotube is larger than the mean free path.
In this case, we can neglect effects of edges on the conductivity of the carbon nanotube and replace the conductivity with
%
\begin{equation}
\sigma(y,y') = \theta_0(y)\theta_0(y')\sigma(y\!-\!y') ,
\end{equation}
%
where $\sigma(y\!-\!y')$ is the conductivity in infinitely long nanotubes and $\theta_0(y)$ is unity well inside the nanotube and should decay rapidly outside.
The simplest choice is
%
\begin{equation}
\theta_0(y) = \theta\Big[\Big({l\over 2}\Big)^2\!-\!y^2\Big] ,
\label{Step Function}
\end{equation}
%
where $\theta(t)$ is the step function defined by
%
\begin{equation}
\theta(t) = \left\{
\begin{array}{cl} 1 & (t>0); \\ 0 & (t<0). \\
\end{array}
\right.
\end{equation}
%
Actually, a nanotube has circumference $L\!=\!2\pi R$ and the field should be smoothed out over the distance of the order of $L$.
Therefore, we shall replace $\theta_0(y)$ with
%
\begin{equation}
\theta_0(y) = {1\over 2} \Big[ 1 \!-\! {\rm erf} \Big( { |y|\!-\!{1\over2} l \over \xi } \Big) \Big] ,
\end{equation}
%
with effective edge width $\xi\!=\!L$, where the error function is defined by
%
\begin{equation}
{\rm erf}(t) = {2\over\sqrt{\pi}} \int_0^t \e^{-t^2} \d t .
\end{equation}
%
When the mean free path is much larger than the tube length, discrete energy levels are formed because of quantization of the electron motion and deviations from this approximation may appear.
\par
%
For simplicity, we consider the case that the Fermi level lies in the linear band
%
\begin{equation}
\varepsilon = \pm \gamma |k| = \pm \hbar v_{\rm F}|k| , 
\end{equation}
%
where $k$ is the electron wave-vector, $v_{\rm F}\!=\!\gamma/\hbar$ is the Fermi velocity, and $\gamma\!=\!\sqrt{3}a\gamma_0/2$ with the lattice constant $a$ and the nearest-neighbor hopping integral $\gamma_0$ in graphene.
Further, we shall employ a relaxation-time approximation to calculate the Boltzmann conductivity, which becomes an exact solution of the transport equation in the single-channel case.
As shown in Appendix A, we have the conductivity
%
\begin{equation}
\sigma(q,\omega) = g_{\rm v}g_{\rm s}{e^2\gamma\over\pi\hbar^2}\Big( \!-\! \i\omega\!+\!{1\over \tau}\Big) \Big[\Big(\!-\!\i\omega\!+\!{1\over\tau}\Big)^2\!+\!(v_{\rm F}q)^2\Big]^{\!-\!1},
\end{equation}
%
where $g_{\rm v}$ and $g_{\rm s}$ are the degeneracy for the valley and spin, respectively.
The impurity and/or phonon scattering are characterized by relaxation time $\tau$, related to mean free path $\Lambda$ given by
%
\begin{equation}
\Lambda\!=\!v_{\rm F} \tau .
\end{equation}
%
\par
%
\begin{figure}
\begin{center}
\leavevmode\includegraphics[width=1.\hsize]{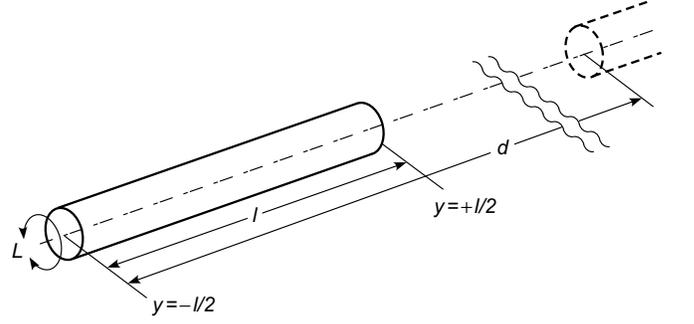}
\end{center}
\caption{
A schematic illustration of a finite-length nanotube with length $l$.
In actual numerical calculations, we shall consider a periodic array with period $d$ ($d\!\gg\!l$).
}
\label{Finite-Length-CN}
\end{figure}
%
\subsection{Periodic Array}
%
For actual numerical calculations, we consider the array with period $d$ as illustrated in Fig.\ \ref{Finite-Length-CN} and seek the solution periodic with $d$.
The electric field is expanded into the Fourier series
%
\begin{equation}
E(y) = \sum_G E(G) \exp(\i G y ) ,
\end{equation}
%
with reciprocal lattice $G\!=\!2\pi j/d$ $(j\!=\!0,\pm1,\cdots)$.
Then the equation for the electric field becomes
%
\begin{equation}
\sum_{G'} \Big[ \delta_{G,G'} - {1 \over \i \omega} {l \over d} K'(G) \sigma(G,G') \Big] E(G') = E_{\rm ext}(G) ,
\label{Equation for Electric Field in Wavevector Space}
\end{equation}
%
where
%
\begin{equation}
\sigma(G,G') = l \int {\d q \over 2\pi} \sigma(q,\omega) \theta_0(G\!-\!q) \theta_0(G' \!-\! q) ^* ,
\end{equation}
%
and
%
\begin{equation}
\theta_0(q) = {1\over l} \int \! \d y \, \theta_0(y) \, \e^{- \i q y}.
\end{equation}
%
In terms of the Fourier coefficients, the power absorption of each nanotube is written as
%
\begin{equation}
P = {l\over 2} {\rm Re} \sum_{G} \sum_{G'} E(G)^* \sigma (G,G') E(G') .
\end{equation}
%
\par
%
Because $K'(q)$ increases in proportion to $q$ for large $q$, the convergence of the solution of Eq.\ (\ref{Equation for Electric Field in Wavevector Space}) requires very large values of $G$.
This slow convergence corresponds to the singular distribution of the electric field associated with point-like polarization charges building up in the vicinity of the end points of the nanotube.
Actually, the use of the nonlocal conductivity is quite effective in suppressing this singular behavior, but is not essential for determining resonance behavior particularly in the low frequency region.
\par
%
\subsection{Single-Mode Approximation}
%
Let us consider the case in a uniform external field, i.e., $E_{\rm ext}(y)\!=\!E_{\rm ext}$, which is valid for light with wavelength much larger than the tube length and polarized in the $y$ direction.
Polarization charge is induced at the ends of the finite length nanotube, as mentioned above.
This charge is regarded almost as a point charge and the associated electric field decreases in proportion to $1/r^2$, where $r$ is the distance from the end point.
In sufficiently long nanotubes, therefore, the field due to this induced charge is negligible in the most region of the nanotube where the absorption mainly takes place.
If we neglect the presence of the ends, a single plasmon mode is induced along the nanotube determined by the wave number corresponding to the frequency of the external field.
For such a long-wavelength mode, the response is essentially determined by local conductivity $\sigma(\omega)\!\equiv\!\sigma(0,\omega)$, i.e., $\sigma(y\!-\!y')\!=\!\sigma(\omega)\delta(y\!-\!y')$.
Further, the main effect caused by the presence of the edges on this mode is to impose the boundary condition that the induced current should vanish at the edges, leading to the vanishing electric field.
\par
%
Let $Q_\omega$ be the wave number of such a mode dominantly excited.
Then, the electric field associated with this ``bulk'' mode is approximately given by
%
\begin{equation}
E(y) \approx E_0 [ \cos (Q_\omega y) \!-\! \cos (Q_\omega l/2) ] ,
\end{equation}
%
with appropriate coefficient $E_0$.
The second term in the bracket arises from the boundary condition, $E(\pm l/2)\!=\!0$.
Approximately, we have
%
\begin{eqnarray}
& &\int \!\! \d y' \int \!\! \d y'' \, K'(y\!-\!y') \sigma(y',y'') E_0 [ \cos (Q_\omega y'') \!-\! \cos (Q_\omega l/2) ] \nonumber\\
& \approx &K'(Q_\omega) \sigma(\omega) E_0 \cos (Q_\omega y) . 
\end{eqnarray}
%
Thus, Eq.\ (\ref{Total Field: Integral Equation}) gives two equations:
%
\begin{eqnarray}
& &1 - {\sigma(\omega) \over \i \omega } K'(Q_\omega) = 0 ,\nonumber\\
& - &E_0 \cos (Q_\omega l/2) = E_{\rm ext} . 
\end{eqnarray}
%
\par
%
The second equation gives
%
\begin{equation}
E(y) \approx E_{\rm ext} \Big( 1 \!-\! {\cos(Q_\omega y) \over \cos(Q_\omega l/2 ) } \Big) .
\label{Single-Mode Approximation}
\end{equation}
%
The first equation gives the mode frequency:
%
\begin{equation}
1 -  {\omega(Q_\omega)^2 \over \omega [ \omega \!+\! (\i/\tau) ] } = 0 ,
\label{Dispersion Relation: Complex}
\end{equation}
%
with
%
\begin{equation}
\omega(Q_\omega)^2 = {g_{\rm v} g_{\rm s} e^2 \gamma \over \pi\hbar^2 } K'(Q_\omega) .
\end{equation}
%
In the limit of an ideal nanotube $\tau\!\rightarrow\!\infty$, we have $Q_\omega\!=\!\pm Q$ with $Q\!>\!0$, satisfying
%
\begin{equation}
\omega(Q)^2 = {g_{\rm v} g_{\rm s} e^2 \gamma\over\pi\hbar^2}{2\over \kappa} K_0 (QR)I_0 (QR) Q^2 = v_Q^2 Q^2 ,
\label{Dispersion Relation}
\end{equation}
%
where $v_Q$ is the approximate velocity of the mode
%
\begin{equation}
v_Q = 4 v_{\rm F} \sqrt{{g_{\rm v} g_{\rm s} \over 4} {e^2 \over 2 \pi\kappa\gamma} K_0 (QR)I_0 (QR) } .
\end{equation}
%
\par
%
\begin{figure}
\begin{center}
\leavevmode\includegraphics[width=1.\hsize]{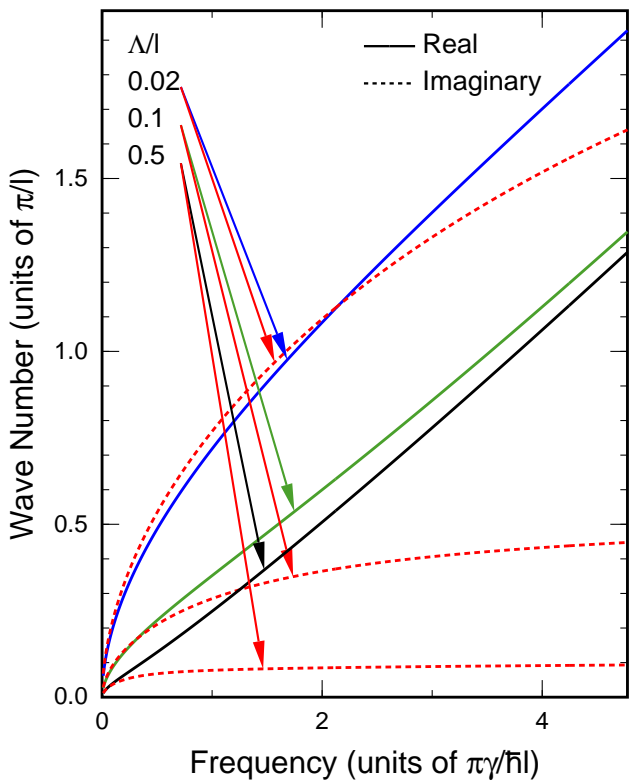}
\end{center}
\caption{
Examples of calculated plasmon dispersion in a tube with radius $R/l\!=\!1/1500$ for $\Lambda/l=0.02$, 0.1, and 0.5, where $\Lambda$ is the mean free path.
Solid and dotted lines show the real and imaginary parts of $Q_\omega$, respectively.
The real part is almost linear with the slope of $v_Q/v_{\rm F} \sim 4$ in the clean tube $\Lambda/l\!=\!0.5$.
}
\label{Dispersion}
\end{figure}
%
Except in extremely short nanotubes with length not so much different from circumference $L$, the velocity is nearly constant because of logarithmically weak dependence on $Q$.
We have $e^2/2\pi\kappa\gamma\,\lsim\,0.2$ and $K_0 (QR)I_0 (QR)\allowbreak\!>\!1$, and therefore $v_Q\!>\!v_{\rm F}$.
In clean wires satisfying $\omega\tau\!\gg\!1$, we have $Q_\omega \!=\! Q \!+\! \i Q'$, with
%
\begin{equation}
Q \approx { \omega \over v_Q }, 
\quad
Q' \approx {1 \over 2 v_Q \tau } = {1\over 2} {v_{\rm F} \over v_Q } {1\over \Lambda} .
\end{equation}
%
In dirty wires $\omega\tau\!\ll\!1$, on the other hand, we have
%
\begin{equation}
Q_\omega \approx {1 \over \sqrt2 v_Q } \sqrt{\omega \over \tau} ( 1 \!+\! \i ) .
\end{equation}
%
\par
%
Figure \ref{Dispersion} shows some examples of $Q_\omega$ for tubes with $l/R\!=\!2\pi l/L\!=\!1500$ (corresponding to a tube with length $l\!\approx\!1$ $\mu$m and diameter $2R\!\approx\!1.36$ nm of the so-called (10,10) nanotube).
We have used $\gamma_0\!=\!3.0$ eV and $\kappa_0\!=\!2.5$ corresponding to the bulk graphite.\cite{Taft and Philipp 1965a}
In this example, we have roughly $v_Q\!\approx\!4v_{\rm F}$.
The velocity is larger than the electron velocity in the graphene or in metallic carbon nanotubes, but is still much smaller than the light velocity $c$ because $v_{\rm F}\!\sim\!c/300$.
This slow plasmon velocity justifies the use of Poisson's equation instead of full Maxwell's equation used for normal metal wires.
The above results become essentially the same as that obtained in ref.\ \onlinecite{Slepyan et al 2006a} based on the so-called Leontovich-Levin equation when we take the limit $c\!\rightarrow\!\infty$.
Plasmon modes have been theoretically studied in infinite one-dimensional organic conductors\cite{Williams and Bloch 1974a} and in semiconductor quantum wires.\cite{Das Sarma and Hwang 1996a}
\par
%
In clean tubes satisfying condition $\cosh(Q'l/2)\allowbreak\!\gg\!\sinh(Q'l/2)$ and therefore $\cosh(Q'y/2)\allowbreak\!\gg\!|\sinh(Q'y/2)|$, we have approximately
%
\begin{eqnarray}
E(y) & \approx &E_{\rm ext} \bigg[ 1 - \cos{Qy\over2} \cosh{Q'y \over 2} \nonumber\\
&\times& \Big( \cos{Ql \over 2} \cosh {Q'l \over 2} \!-\! \i \sin{Ql \over 2} \sinh{Q'l \over 2} \Big)^{-1} \bigg] , 
\end{eqnarray}
%
except in narrow regions of $y$ for which $\cos(Qy/2)\!\approx\!0$.
This exhibits a resonance behavior at $Q\!=\!Q_n$ with
%
\begin{equation}
Q_n \equiv (2n \!+\! 1) {\pi \over l} \quad
(n\!=\!0,1,\cdots) .
\end{equation}
%
In fact, in the vicinity of $Q_n$, we have
%
\begin{eqnarray}
E(y) & \approx &E_{\rm ext} (-1)^n \Big( \! \cos{Q'l \over 2}\Big)^{-1} \cos{Qy\over2} \cosh{Q'y \over 2} \nonumber\\
& \times &\Big[ {(Q\!-\!Q_n)l \over 2} \!+\! \i \tanh {Q'l \over 2} \Big]^{-1} ,
\end{eqnarray}
%
showing that the imaginary part exhibits a resonant increase at $Q_n$ following a Lorentzian form and the real part a Lorentzian multiplied by $Q\!-\!Q_n$.
Therefore, the absorption also exhibits a resonance proportional to
%
\begin{eqnarray}
P & \propto &N l E_{\rm ext}^2 {\sigma_0\over 1+(\omega\tau)^2}
\Big( \! \cos{Q'l \over 2}\Big)^{-2}\nonumber\\
& \times & \Big[ {(Q\!-\!Q_n)^2l^2 \over 4} \!+\! \tanh^2 {Q'l \over 2} \Big]^{-1} , 
\end{eqnarray}
%
with $\sigma_0\equiv \sigma(0,0)=g_v g_s (e^2/\pi\hbar)\Lambda$ and total number $N$ of isolated nanotubes.
\par
%
This resonance behavior decreases with the increase of disorder or the decrease of mean free path $\Lambda$.
The clear resonance disappears when $\cosh(Q'l/2)\!\approx\!\sinh(Q'l/2)$, i.e., $Q'\!\sim\!\pi/l$.
In the present example with $v_Q\!\sim\!4v_{\rm F}$, this condition becomes $\Lambda/l\!\sim\!0.04$.
In the dirty limit, the field remains small for small $\omega$ or $Q$ ($\approx Q'$) until condition $Q\!\sim\!Q'\!\sim\!\pi/l$ is satisfied and then becomes essentially the same as the external field except in the vicinity of the edges.
As a result, the absorption takes a broad maximum around the corresponding frequency and decays with $\omega$ following the real part of the dynamical conductivity $\sigma(\omega)$.
\par
%
\begin{figure}
\begin{center}
\leavevmode\includegraphics[width=1.\hsize]{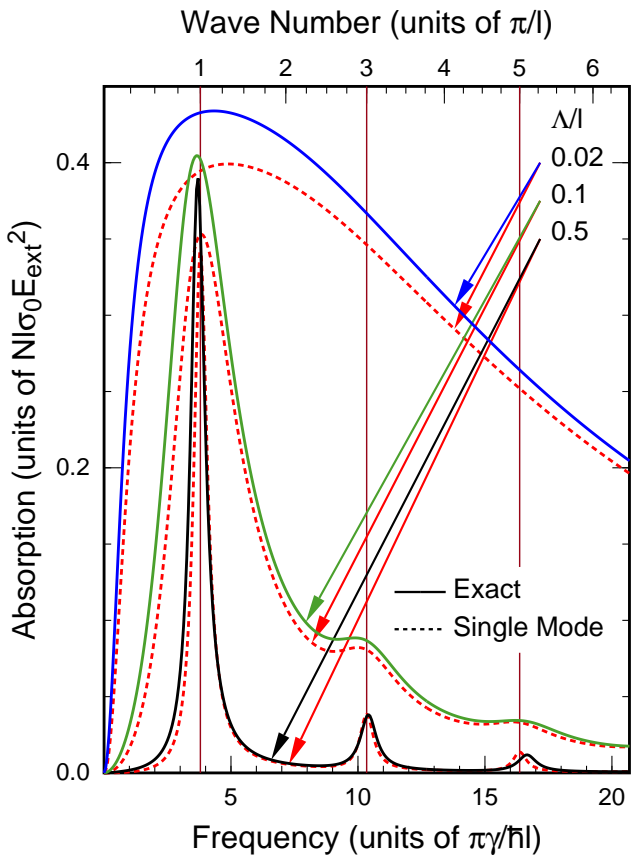}
\end{center}
\caption{
The absorption in a nanotube with radius $R/l\!=\!1/1500$ as a function of frequency.
$\Lambda/l\!=\!0.5$, 0.1, and 0.02.
The results in the single-mode approximation are shown by dotted lines.
Corresponding wave number $Q_\omega$ in an infinitely long tube with $\Lambda\!\rightarrow\!\infty$ is shown in the upper axis.
Thin vertical dashed lines indicates wave number $Q_n\!=\!\pi(2n\!+\!1)/l$.
}
\label{Absorption}
\end{figure}
%
\begin{figure}
\begin{center}
\leavevmode\includegraphics[width=1.\hsize]{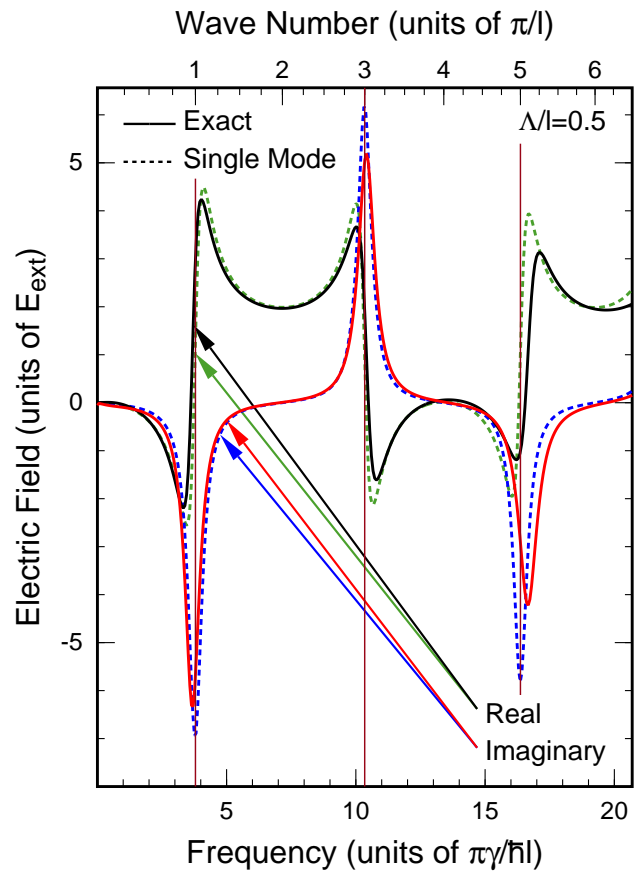}
\end{center}
\caption{
The real and imaginary parts of the electric field at the center of the nanotube as a function of frequency.
The dotted lines represent those of the single-mode approximation.
}
\label{Electric Field at Tube Center}
\end{figure}
%
\section{Numerical Results}
%
\begin{figure*}
\begin{center}
\leavevmode\includegraphics[width=0.75\hsize]{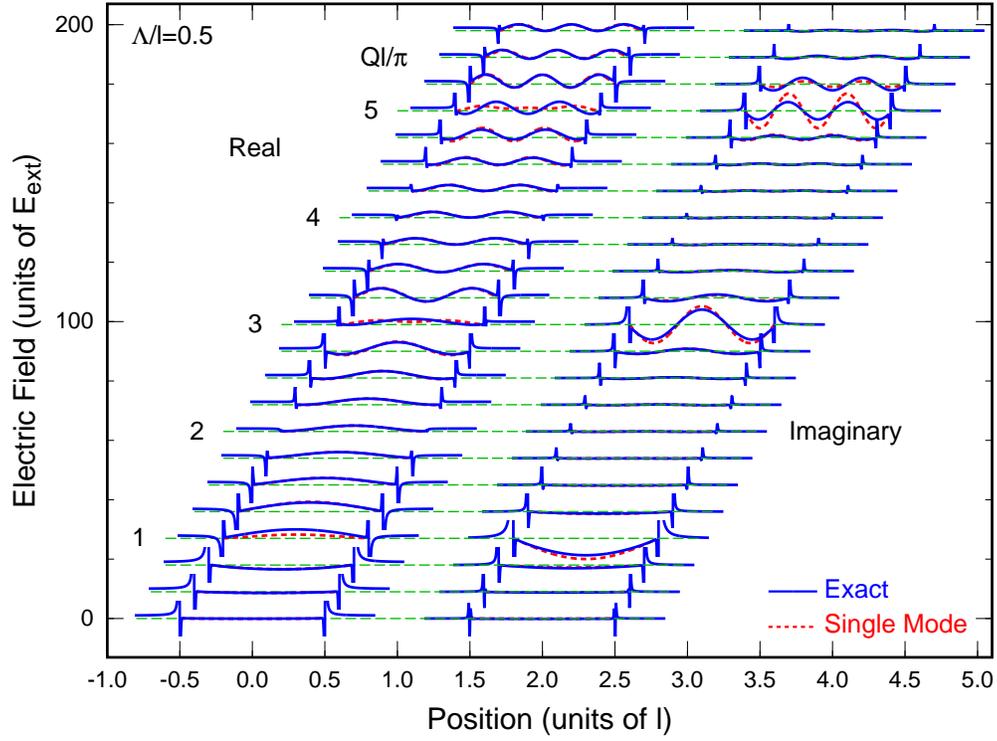}
\end{center}
\caption{
Calculated electric field distribution for varying $Q$ in a clean wire with mean free path $\Lambda/l=0.5$.
The dotted lines represent the results of the single-mode approximation.
Lines for increasing $Ql/\pi$ are shifted in the vertical direction, for each of which the horizontal axis at $E=0$ is shown by a dashed line.
}
\label{Electric Field: 0.5}
\end{figure*}
%
\begin{figure*}
\begin{center}
\leavevmode\includegraphics[width=0.75 \hsize]{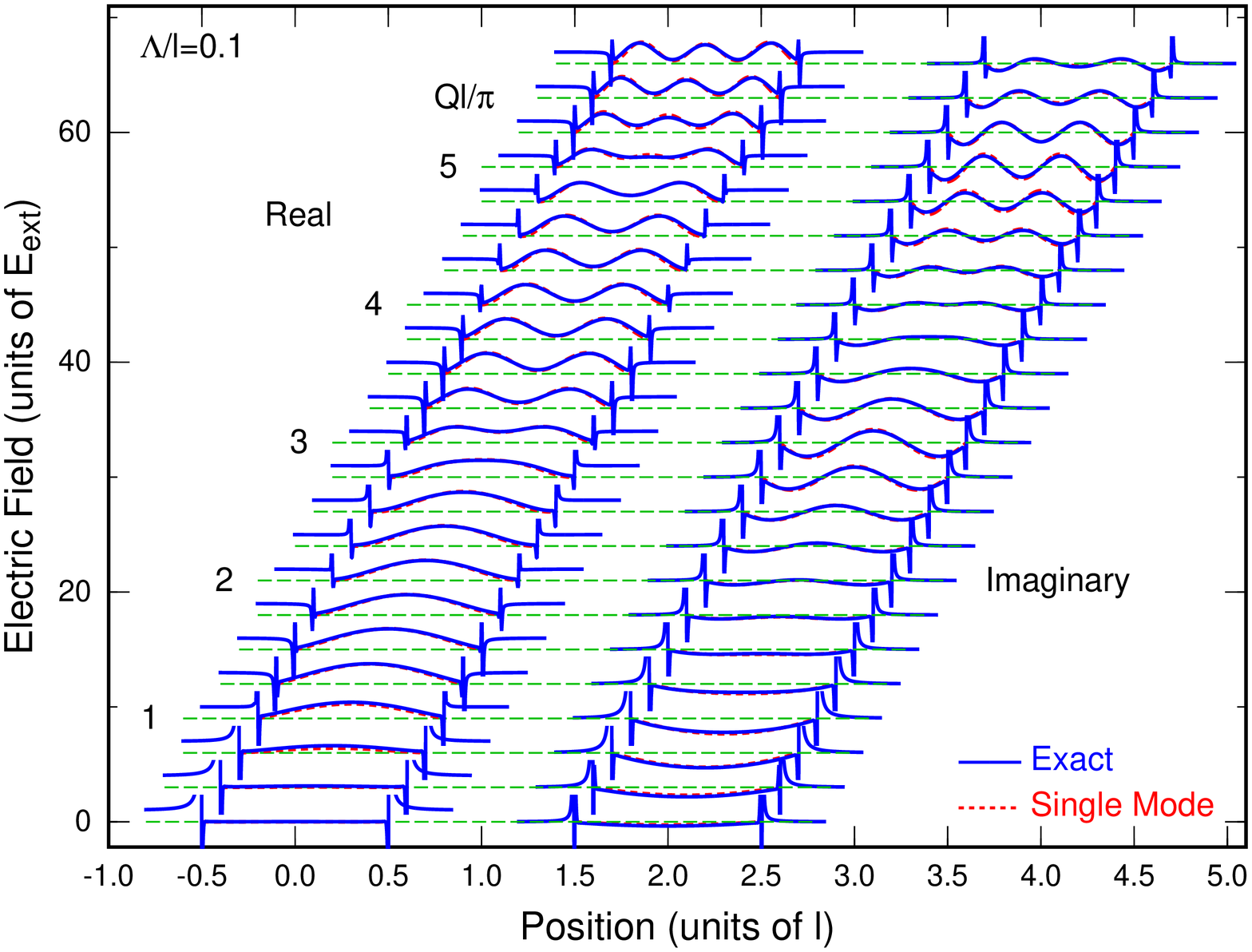}
\end{center}
\caption{
Calculated electric field distribution in a wire with $\Lambda/l=0.1$.
}
\label{Electric Field: 0.1}
\end{figure*}
%
Figure \ref{Absorption} shows some examples of calculated absorption intensity in the nanotube the same as shown in Fig.\ \ref{Dispersion}.
The mean free path is chosen to be $\Lambda/l\!=\!0.5$ (clean), 0.1 (intermediate), and 0.02 (dirty) in these examples.
The solid lines show numerical results and the dotted lines those obtained in the single-mode approximation discussed in the previous section.
The positions corresponding to $Q\!=\!Q_n$ are denoted by thin vertical lines.
\par
%
The main resonance with $n\!=\!0$ occurs almost exactly at $Q\!=\!Q_0$.
For the higher-frequency resonance $n\!=\!1$, the peak position is slightly shifted to the higher frequency side.
This shift becomes larger with the increase of $n$.
In the single-mode approximation, on the other hand, the resonance occurs exactly at $Q_n$.
This shift is due to the appearance of strongly localized electric field associated with induced charges at both ends of the nanotube.
Apart from this peak shift, the single-mode approximation works surprisingly well.
\par
%
Figure \ref{Electric Field at Tube Center} shows the electric field at the center, $y\!=\!0$, of the clean tube ($\Lambda/l\!=\!0.5$) as a function of the frequency.
The real and imaginary parts exhibit the resonance behavior same as that discussed in the previous section within the single-mode approximation.
In fact, the single-mode approximation has originally been expected to be sufficient for the behavior of the field in the central part of the nanotube.
\par
%
Figure \ref{Electric Field: 0.5} shows calculated electric-field distribution in a clean tube with $\Lambda/l\!=\!0.5$.
The dotted lines show the results in the single-mode approximation.
The most noteworthy feature is large electric field localized at both ends of the nanotube.
This is due to the significant accumulation of induced charges at both ends of the tube.
This field decays rapidly away from the tube edge and approaches external field $E_{\rm ext}$ outside the tube.
Well inside the tube, the field becomes approximately sinusoidal with wave vector $Q$ of the plasmon mode determined by the frequency.
In the low-frequency region corresponding to $Q$ smaller than $Q_0$, the electric field is screened out in a carbon nanotube by charges accumulated at both ends.
At resonances, both real and imaginary parts of the field are considerably enhanced.
\par
%
Difference between the exact numerical results and those of the single-mode approximation becomes apparent in the vicinity of resonance $n\!=\!2$, i.e., $Ql/\pi\!=\!5$.
This arises mainly due to the shift in the peak frequency in the numerical result, not present in the single-mode approximation.
For larger $n$, the field distribution is more strongly affected by accumulated edge charges and extra restoring force tends to enhance the resonance frequency from that in the single-mode approximation.
\par
\begin{figure*}
\begin{center}
\leavevmode\includegraphics[width=0.75\hsize]{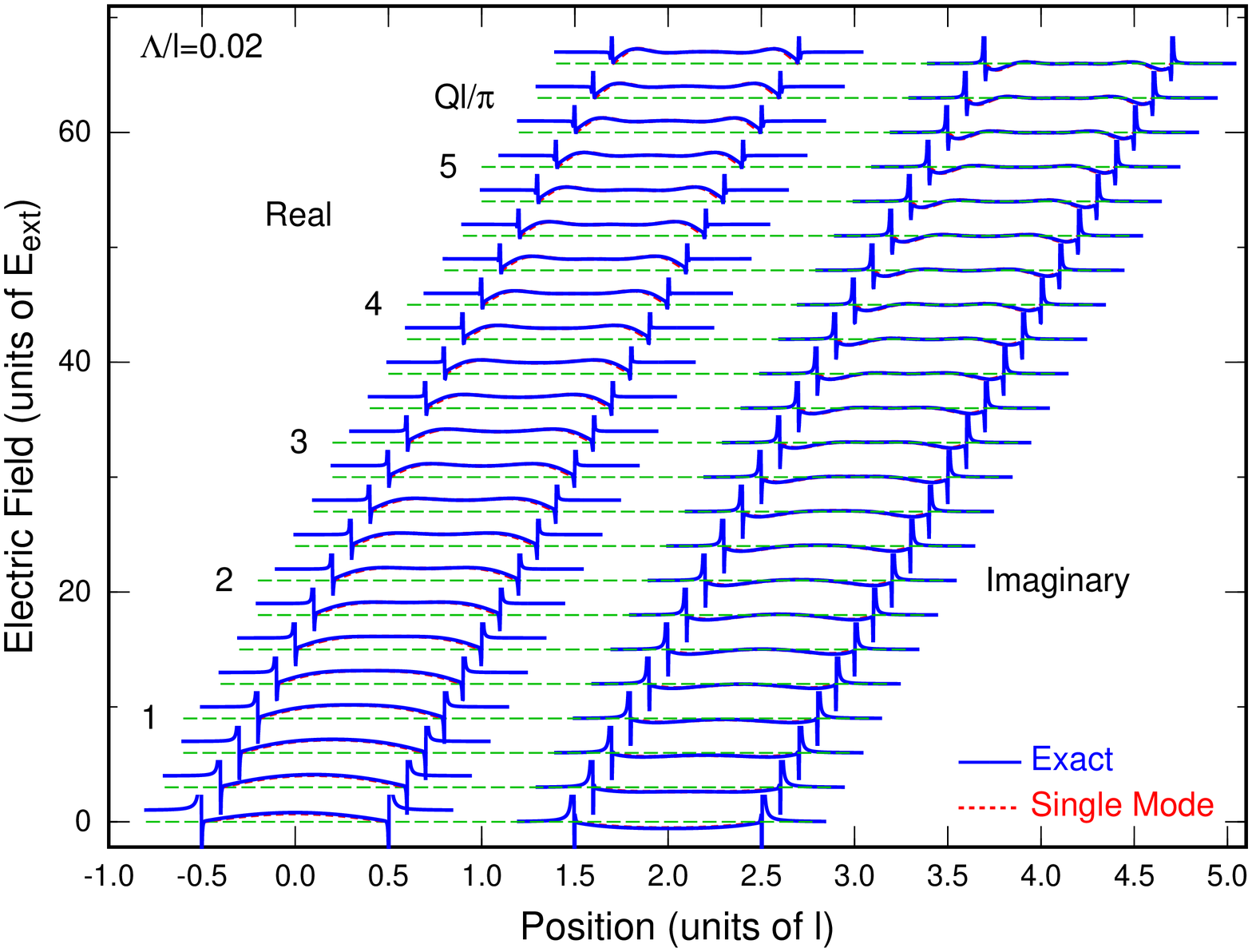}
\end{center}
\caption{
Calculated electric field distribution in a dirty wire with $\Lambda/l=0.02$.
}
\label{Electric Field: 0.02}
\end{figure*}

%
Figure \ref{Electric Field: 0.1} shows calculated electric-field distribution in a tube with $\Lambda/l\!=\!0.1$, for which $Q'l\!\sim\!1$.
In this case broad resonance is recognized only in the imaginary part of the electric field and deviation from the single-mode results is less prominent.
In a dirty tube with $\Lambda/l\!=\!0.02$ shown in Fig.\ \ref{Electric Field: 0.02}, the field distribution is almost independent of the frequency except in the low frequency region where $Q\!\sim\!Q'\,\lsim\,\pi/l$, as has been discussed in the previous section for the results in the single-mode approximation.
\par
%
\begin{figure}
\begin{center}
\leavevmode\includegraphics[width=1.\hsize]{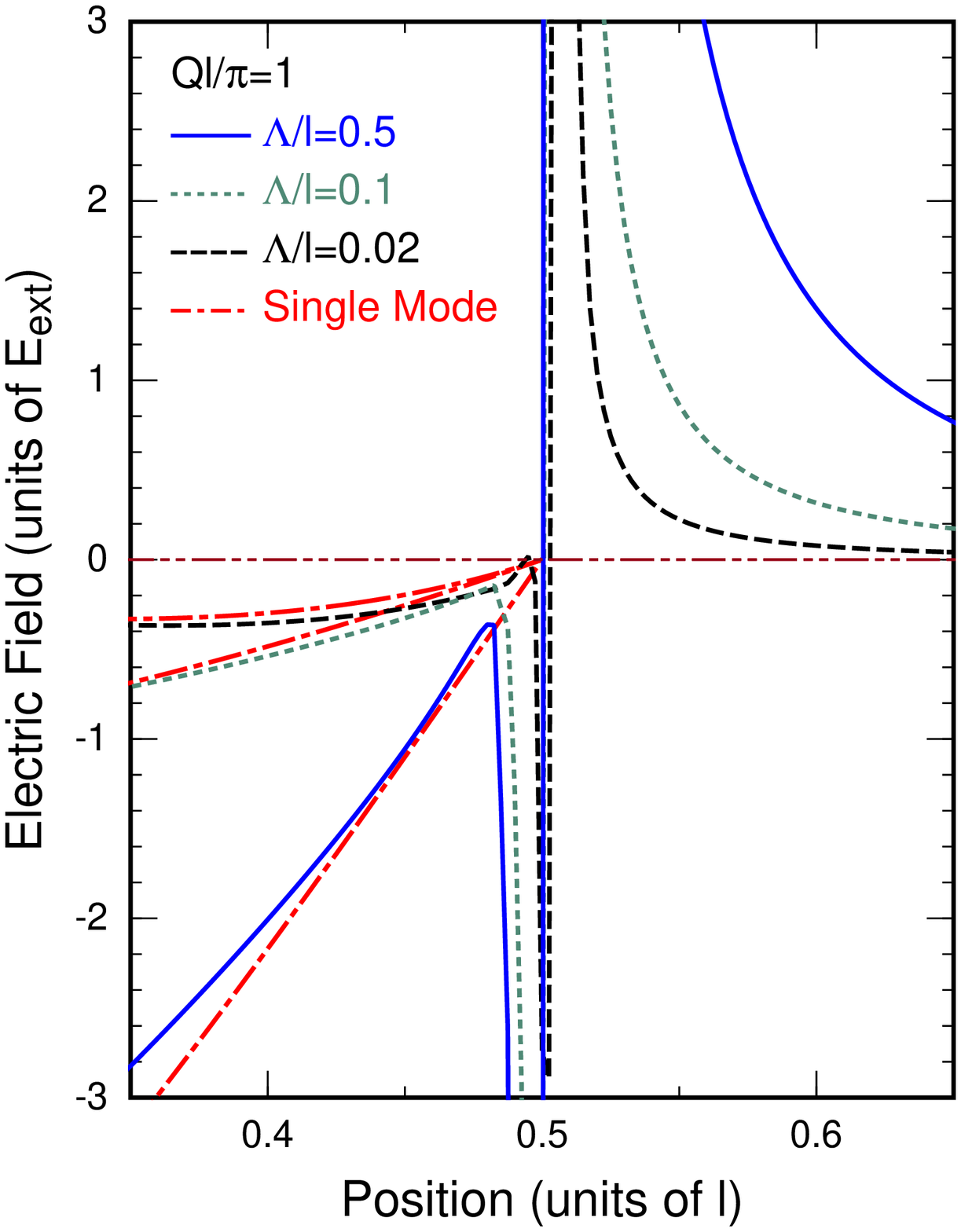}
\leavevmode\includegraphics[width=1.\hsize]{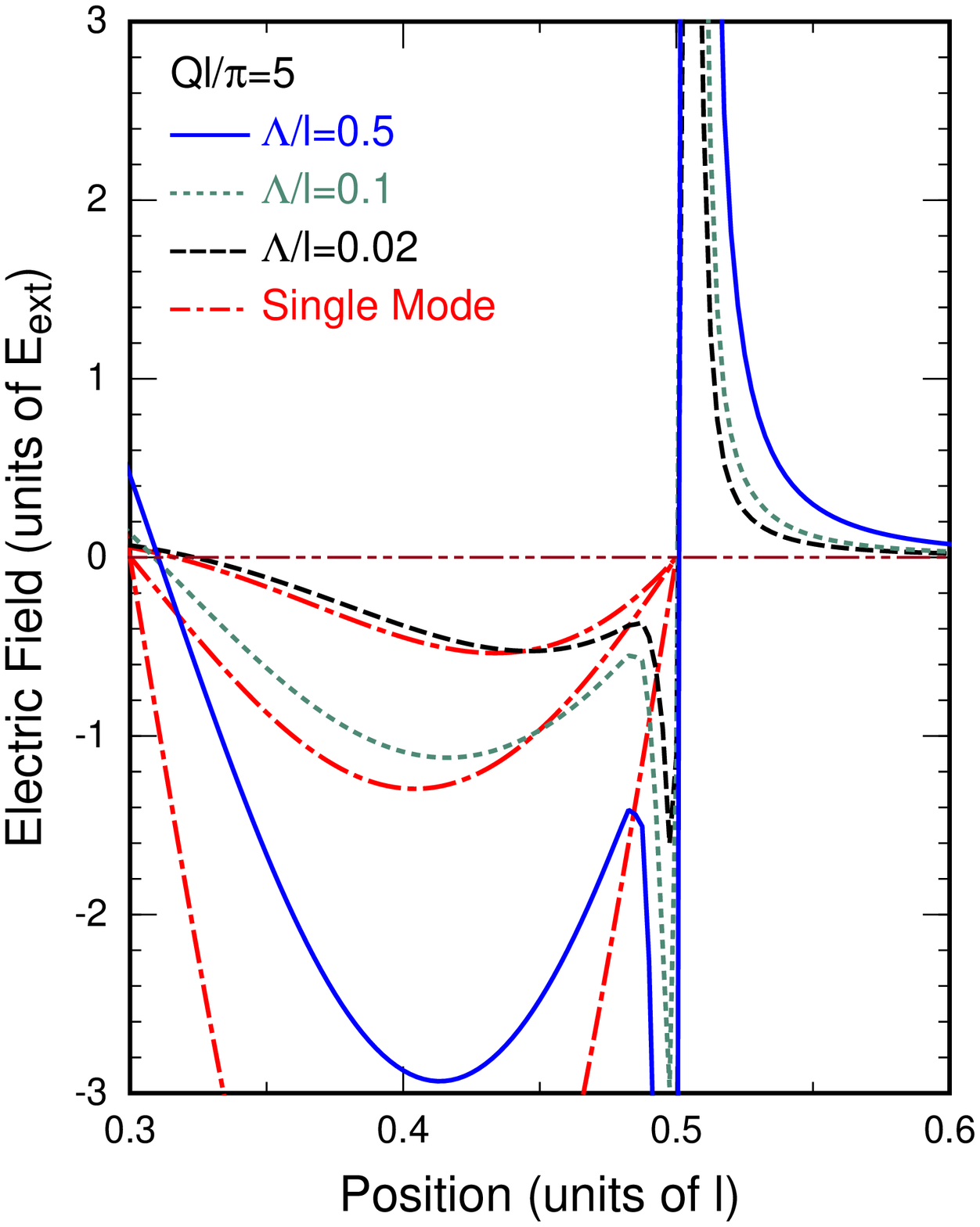}
\end{center}
\caption{
A close-up view of imaginary part of electric field distribution around the right end of nanotubes for (a) $Ql/\pi\!=\!1$ and (b) $5$.
}
\label{Ends}
\end{figure}

%
Figure \ref{Ends} shows the details of field distribution (imaginary part of $E(y)$) around the right end of the nanotube for (a) $Ql/\pi\!=\!1$ and (b) $5$.
Polarized charges appear at the edge roughly in proportion to the conductivity and therefore their amount decreases with the decrease of the mean free path.
A large electric field appears associated with these charges and decays slowly in proportion to $(y\!-\!{1\over2}l)^{-2}$ outside of the nanotube.
This outside field does not contribute to the power absorption, however.
\par
%
%
On the other hand, the electric field due to edge charges rapidly decays inside the nanotube due to strong screening effect and becomes negligible in the most region of the nanotube where the absorption mainly takes place.
In fact, only small difference appears in the absorption power given by the solid and dotted lines in Fig.\ \ref{Absorption}.
This shows that the strong screening of the field due to polarized edge charges is the main reason that the single-mode approximation based on a plasmon mode in an infinitely long nanotube has turned out to work surprisingly well in spite of the presence of edges.
\par
%
\section{Discussion and Conclusion}
%
Carbon nanotubes used in experiments\cite{Akima et al 2006a} are usually several micrometer long and therefore longer than the mean free path limited by impurity and/or phonon scattering at room temperature.
Recent progress in experiments has achieved short nanotubes with sub-micron size.\cite{Ziegler et al 2005a,Ziegler et al 2005b,Arnold et al 2006a,Hennrich et al 2007a,Fagan et al 2008a,Sun et al 2008a}
In such short nanotubes, discrete energy levels are formed and the level spacing can exceed the broadening due to disorder.
Then, we should seriously consider effects of discrete energy levels to discuss optical absorption of finite-length nanotubes.
This problem is out of the scope of this paper and left for a future study.
\par
%
When a nanotube is tilted from the direction of the external field, the field component parallel to the axis is effective in the absorption and the resonance frequency remains unaffected.
When many nanotubes are distributed at random within a plain, the absorption intensity is reduced by factor $1/2$ due to the average over directions.
Recently, a monotonic dependence on the direction is reported in a mat of aligned nanotubes.\cite{Ren et al 2009a}
For the field perpendicular to the axis, there is no significant absorption in the relevant frequency range considered here.\cite{Benedict et al 1995a,Novikov and Levitov 2006a,Yamamoto et al 2008a}
In fact, the dynamical screening or the depolarization effect due to induced charges should be considered\cite{Ajiki and Ando 1994a,Ajiki and Ando 1995d} and resonance absorption appears at much higher energies when exciton effects are properly included in interband optical transitions.\cite{Uryu and Ando 2006c,Uryu and Ando 2007c}
\par
%
In actual absorption experiments, a bundle of nanotubes may be used.
When a bundle contains $N$ metallic nanotubes with same length, the Coulomb kernel given by Eq.\ (\ref{Kernel}) is effectively multiplied by $N$.
As a result, the frequency and the velocity of excited plasmon mode are multiplied by $\sqrt{N}$.
Semiconducting nanotubes can be contained in a bundle, but usually have very small conductivity and irrelevant, although it is known that carbon nanotubes tend to be naturally doped due to surrounding particles such as oxygen.\cite{Jhi et al 2000a}
\par
%
Usually, experiments are performed for a film where the density of nanotubes is sufficiently small for the purpose of avoiding electrical contacts among them.
In this case, electric field induced by neighboring nanotubes can be considered in a dipole approximation.
The dipole field decays as $1/r^3$ with the increase in distance $r$.
In film-like systems, this dipole field of distant nanotubes does not give rise to significant contribution in contrast to three-dimensional systems.
Effects of surrounding nanotubes were previously considered by numerically solving Hall\'en's equation for infinite planar arrays of nanotubes.\cite{Hanson 2005a,Hao and Hanson 2006a}
The results showed that interaction between neighboring tubes causes broadening and shift of resonances only when the distance becomes of the order of the tube diameter.
If an end of a nanotube touches another nanotube, the absorption may be significantly modified because of large charges accumulated at the end.
Such a situation should be avoided in experiments.
\par
%
In general, a mini-gap opens around the Fermi energy even in so-called metallic nanotubes for various reasons.
For example, the nonzero curvature causes a band gap depending on the chirality,\cite{Hamada et al 1992a} which can be understood in terms of an effective Aharonov-Bohm flux within the {\bf k}$\cdot${\bf p} scheme.\cite{Ando 2000b,Kane and Mele 1997a,Ando 2005a}
A mechanical deformation causes band-gap modification,\cite{Yang et al 1999a,Yang and Han 2000a} which can also be understood in terms of flux.\cite{Kane and Mele 1997a,Suzuura and Ando 2002a}
The amount of the gap depends on the chirality, radius, environment, etc., and is typically smaller than room temperature.
There have been some reports suggesting the observation of broad absorption due to inter-minigap transitions.\cite{Itkis et al 2002a,Kampfrath et al 2008a}
Because this mini-gap transition is in the same frequency region, detailed study on the length dependence is required for the purpose of identifying the absorption due to the finite-length origin considered here.
\par
%
In summary, we have calculated electric-field distribution and absorption intensity of a finite-length nanotube in oscillating electric field.
The results show that the main resonance corresponding to excitation of the fundamental plasmon mode with wave vector $Q_0\!=\!\pi/l$ is quite robust except in very dirty tubes.
For higher-frequency resonances, the electric field associated with induced edge charges starts to be mixed and tends to shift resonances to higher frequencies.
Overall resonance behaviors can be reasonably well described by the single-mode approximation in which effects of induced edge charges are completely neglected.
\par
%
\section*{ACKNOWLEDGMENTS}
%
We thank fruitful discussions with S. Iijima, Y. Iwasa, S. Ohmori, T. Okazaki, and T. Saito.
This work was supported in part by Grant-in-Aid for Scientific Research on Priority Area ``Carbon Nanotube Nanoelectronics,'' by Grant-in-Aid for Scientific Research, by Global Center of Excellence Program at Tokyo Tech ``Nanoscience and Quantum Physics'' from Ministry of Education, Culture, Sports, Science and Technology Japan.
\par
%
\appendix
\section{Nonlocal Conductivity}
%
We start with a Boltzmann transport equation
%
\begin{eqnarray}
& &{\partial f_{jk}(y,t) \over \partial t} + v_{jk} {\partial f_{jk}(y,t) \over \partial y} - e E(y,t) {1\over \hbar} {\partial f_{jk}(y,t) \over \partial k} \nonumber\\
& = &- \sum_{j'k'} W_{j'k',jk} [ f_{jk}(y,t) \!-\! f_{j'k'}(y,t) ] , 
\end{eqnarray}
%
where $j$ denotes subbands, $k$ the wave vector in the axis ($y$) direction, $E(y,t)$ is the applied electric field, $v_{jk}\!=\!\partial\varepsilon_{jk}/\hbar\partial k$ is the velocity, and $W_{j'k',jk}$ is the scattering probability between states $jk$ and $j'k'$.
We write the distribution function $f_{jk}$ as the sum of the equilibrium distribution function $f(\varepsilon_{jk})$ and the deviation due to the applied field $g_{jk}$, i.e., $f_{jk}(y,t) \!=\! f(\varepsilon_{jk}) \!+\! g_{jk}(y,t)$.
In the limit of a weak applied field, the above is approximated by
%
\begin{eqnarray}
 & &{\partial g_{jk}(y,t) \over \partial t} + v_{jk} {\partial g_{jk}(y,t) \over \partial y} + e v_{jk} E(y,t) \Big( \!-\! {\partial f(\varepsilon_{jk}) \over \partial \varepsilon_{jk} } \Big) \nonumber\\
& =& - \sum_{j'k'} W_{j'k',jk} [ g_{jk}(y,t) \!-\! g_{j'k'}(y,t) ] . 
\end{eqnarray}
%
Let
%
\begin{equation}
E(y,t) = E \exp( - \i \omega t \!+\! \i q y ) .
\end{equation}
%
Then, we can set
%
\begin{equation}
g_{jk}(y,t) = g_{jk} \exp( - \i \omega t \!+\! \i q y ) ,
\end{equation}
%
and rewrite the transport equation as
%
\begin{eqnarray}
&  &[ - \i \omega \!+\! \i q v_{jk} ] g_{jk} + e v_{jk} E \Big( \!-\! {\partial f(\varepsilon_{jk}) \over \partial \varepsilon_{jk} } \Big) \nonumber\\
& = &- \sum_{j'k'} W_{j'k',jk} [ g_{jk} \!-\! g_{j'k'} ] . 
\end{eqnarray}
%
\par
%
The solution can be written as
%
\begin{equation}
g_{jk} = \tilde g_{jk} \Big( \!-\! {\partial f(\varepsilon_{jk}) \over \partial \varepsilon_{jk} } \Big) ,
\end{equation}
%
with
%
\begin{equation}
[ - \i \omega \!+\! \i q v_{jk} ] \tilde g_{jk} + e v_{jk} E = - \sum_{j'k'} W_{j'k',jk} [ \tilde g_{jk} \!-\! \tilde g_{j'k'} ] .
\end{equation}
%
This constitutes a set of linear equations determining $\tilde g_{jk}$ for each energy and therefore can be solved exactly.
In the following, however, we shall use a simplest relaxation-time approximation, which become exact for a single channel case.
\par
%
The right hand side is approximated by
%
\begin{equation}
- \sum_{j'k'} W_{j'k',jk} [ g_{jk}(y,t) \!-\! g_{j'k'}(y,t) ] = - {1\over \tau(\varepsilon_{jk})} g_{jk}(y,t) .
\end{equation}
%
Then, we immediately have
%
\begin{equation}
g_{jk} = - \Big[ - \i \omega \!+\! \i q v_{jk} \!+\! {1\over \tau(\varepsilon_{jk})} \Big]^{-1} e v_{jk} E \Big( \!-\! {\partial f(\varepsilon_{jk}) \over \partial \varepsilon_{jk} } \Big) .
\end{equation}
%
The induced current 
%
\begin{equation}
j(q,\omega) = g_{\rm v} g_{\rm s} \sum_{jk} (-e) v_{jk} g_{jk},
\end{equation}
%
can be expressed in terms of the conductivity $\sigma(q,\omega)$ as
%
\begin{equation}
j(q,\omega) = \sigma(q,\omega) E ,
\end{equation}
%
with
%
\begin{equation}
\sigma(q,\omega) = \int \Big( \!-\! {\partial f(\varepsilon) \over \partial \varepsilon} \Big) \sigma(q,\omega,\varepsilon) \d \varepsilon ,
\end{equation}
%
and
%
\begin{equation}
\sigma(q,\omega,\varepsilon) = g_{\rm v} g_{\rm s} \sum_{jk} \Big[ - \i \omega \!+\! \i q v_{jk} \!+\! {1\over \tau(\varepsilon)} \Big]^{-1} e^2 v_{jk}^2 \delta(\varepsilon\!-\!\varepsilon_{jk}) .
\end{equation}
%
Adding the contributions of positive and negative values of $k$, we can rewrite the above as
%
\begin{eqnarray}
\sigma(q,\omega,\varepsilon) = g_{\rm v} g_{\rm s} \sum_{jk\!>\!0}& & e^2 |v_{jk}|^2 \Big[ \Big( \!-\! \i \omega \!+\! {1\over \tau(\varepsilon)} \Big)^2 \!+\! q^2 |v_{jk}|^2 \Big]^{-1} \nonumber\\
& \times & \Big[ \!-\! \i \omega  \!+\! {1\over \tau(\varepsilon)} \Big] \delta(\varepsilon\!-\!\varepsilon_{jk}) . 
\end{eqnarray}
%
\par
%
We shall consider the energy region within the linear band.
Substituting the explicit expression $v_{k}\!=\!v_{\rm F}\!=\!\gamma/\hbar$ and $\varepsilon_k\!=\!\gamma k$, we have
%
\begin{equation}
\sigma(q,\omega,\varepsilon) = g_{\rm v} g_{\rm s} {e^2 \gamma \over \pi\hbar^2 } \Big[ \!-\! \i \omega  \!+\! {1\over \tau} \Big] \Big[ \Big( \!-\! \i \omega \!+\! {1\over \tau} \Big)^2 \!+\! { \gamma^2 q^2 \over\hbar^2 } \Big]^{-1} ,
\end{equation}
%
with $\tau\!=\!\tau(\varepsilon)$ for simplicity.
\par
%
The poles of the conductivity are given by
%
\begin{equation}
\hbar\omega = \pm \gamma q - \i \, { \hbar \over \tau } ,
\end{equation}
%
which corresponds to the subband dispersion in the limit of $\tau\!\rightarrow\!\infty$, as is expected.
In the limit of small $1/\tau$ and $\omega$, the denominator is a function of $Dq^2\!-\!\i\omega$ as is expected, where the diffusion constant is given by
%
\begin{equation}
D = {1\over 2} {\Lambda^2 \over \tau } = {1\over 2} v_{\rm F}^2 \tau.
\end{equation}
%
\par
%
In the high-frequency limit $\omega\tau\!\gg\!1$, on the other hand, the cut-off wave vector $q_{\rm c}$, where the conductivity deviates from the local conductivity $\sigma(\omega)\!\equiv\!\sigma(0,\omega)$,  becomes $v_{\rm F}q_{\rm c}\!\approx\!\omega$ or
%
\begin{equation}
{q_{\rm c} L \over 2\pi} \approx \hbar \omega \Big( {2\pi\gamma \over L}\Big)^{-1} ,
\end{equation}
%
which should be much smaller than unity under the usual conditions.
\par
%

\end{document}